\documentclass[aps,prb,amsmath,twocolumn,showpacs]{revtex4}
\usepackage{graphicx}
\usepackage{dcolumn}
\usepackage{bm}
\usepackage{color}

\begin{document}
\title{Quantum phonon transport of molecular junctions amide-linked with carbon nanotubes: a first-principle study}
\author{J. T. L\"u}
\email{tower.lu@gmail.com}
\author{Jian-Sheng Wang}
\affiliation{
Center for Computational Science and Engineering and Department of Physics,
National University of Singapore, Singapore 117542, Republic of Singapore
}
\date{8 July 2008}
\begin{abstract} 
Quantum phonon transport through benzene and alkane chains amide-linked with
single wall carbon nanotubes (SWCNTs) is studied within the level of density functional
theory. The force constant matrices are obtained from standard quantum
chemistry software. The phonon transmission and thermal conductance are from
the nonequilibrium Green's function and the mode-matching method. We find that
the ballistic thermal conductance is not sensitive to the compression or
stretching of the molecular junction. The terminating groups of the SWCNTs at the
cutting edges only influence the thermal conductance quantitatively. The
conductance of the benzene and alkane chains shows large difference.  Analysis
of the transmission spectrum shows that (i) the low temperature thermal
conductance is mainly contributed by the SWCNT transverse acoustic modes, (ii)
the degenerate phonon modes show different transmission probability due to the
presence of molecular junction, (iii) the SWCNT twisting mode can hardly be
transmitted by the alkane chain. As a result, the ballistic thermal conductance
of alkane chains is larger than that of benzene chains below $38$ K, while it
is smaller at higher temperature.
\end{abstract}
\pacs{05.60.Gg, 44.10.+i, 63.22.-m, 65.80.+n}
\maketitle

\section{Introduction}
The electronic transport properties of a molecular device depend much on the
underlying nuclear configuration and the electron-phonon coupling\cite{galperin-review}.
The mutual interaction between them induces Joule heating in the molecule,
which may have important consequences on the functionality and reliability of
a molecular device. Study of the heat transport by phonons in molecular
junctions is crucial for a better understanding of their electronic
transport properties. Phonon thermal transport itself is also interesting.
Detailed understanding of the underlying transport mechanism is especially
useful for the design of novel phononic
devices\cite{li:184301,science-thermal}. Furthermore, combined study of the
electronic and phononic transport in molecular junctions is the first step
toward the design of molecular thermoelectric
devices\cite{reddy07}.  Great experimental progress has been
made in these directions recently, which enables one to measure the thermal
and thermoelectric transport properties of molecular
junctions\cite{wangzh07,reddy07}. Advances in experimental technique call for
new theoretical method to predict the thermal conductance of molecular
junctions. Although semiempirical or minimal model calculation is helpful to
understand the underlying physics\cite{segal:6840,buldum99}.  For a detailed
quantitative study, a parameter-free, first-principle method is highly
desirable\cite{mingo08}. Furthermore, for many molecular structures there
exists no empirical inter-atom potential.

In this paper we introduce such a method based on a standard quantum chemistry
software, the Gaussian03 parckage\cite{gaussian03}.  Given the molecule
structure, we can obtain the force constant matrices after performing the
energy minimization. The thermal conductance can be calculated via available
theoretical methods. Among them are the nonequilibrium Green's function (NEGF)
method\cite{wang:033408,wang-2007,wang-review-epjb,mingo06,yamamoto06}, which
has been successfully used to  predict the electronic conductance of molecular
junctions\cite{taylor01,brandbyge02,damle01}. To study the transport of each
phonon mode, we will also use the mode-matching method\cite{ando91}, which is
equivalent to the NEGF method in the ballistic limit\cite{khomyakov05}.  Using
these methods, we first compare the thermal conductance of a benzene ring
amide-linked with two $(6,0)$ single-walled carbon nanotubes (SWCNTs) under
different compression or stretching conditions.  Then we study the effect of
SWCNT terminating group at the cutting edges. Finally, we compare the phonon
transmission probability and the thermal conductance of benzene and alkane
chains with the same leads. Although the electron contribution to the thermal
conductance may be comparable with or larger than that of phonons, inclusion of
this effect is out of the scope of present study.

\section{Molecular structure and theoretical method}
\label{sec:theory}
In this section we first introduce the system Hamiltonian and the procedure to
obtain it from the Gaussian03 package. Then we briefly outline the NEGF, the
mode-matching method, and their relationship. For a detailed discussion, we
refer the reader to Ref.~\onlinecite{wang-review-epjb}. The system we are
interested in is a molecular junction connected with two periodic semi-infinite
leads at both sides. It is a standard treatment to divide the whole structure
into three parts: the center junction and two leads acting as thermal baths.
The boundaries between the center and the baths may be at arbitrary positions,
and not correspond to any physical interface. But it is desirable to include
part of the leads into the center region, since we need to make sure that there
is no direct interaction between the two baths, which is required by the NEGF
formalism. By doing this, we can also include the charge transfer effect
between the leads and the molecule junction. In this setup, the system
Hamiltonian can be written as
\begin{equation}
\label{eq:h}
{\cal H} = \!\!\!\!\!\sum_{\alpha=L,C,R}\!\!\!\!\!H_\alpha  + (u^L)^T V^{LC} u^C + (u^C)^TV^{CR} u^R + V_n,
\end{equation}
where $H_{\alpha} = \frac{1}{2} {(\dot{u}^\alpha)}^T \dot{u}^\alpha +
\frac{1}{2} {(u^\alpha)}^T K^\alpha u^\alpha$ represents harmonic
oscillators, $u^\alpha$ is a column vector consisting of all the mass normalised
displacement variables in region $\alpha$, and $\dot{u}^\alpha$ is the
corresponding conjugate momentum.  $K^\alpha$ is the spring constant matrix in the tight-binding form
and $V^{LC}=(V^{CL})^T$ is the coupling matrix of the left lead to the
central region; similarly for $V^{CR}$. $V_n$ is the nonlinear interaction
in the center, which could be $V_n = \frac{1}{3}\sum_{ijk}t_{ijk}u_iu_ju_k$
for cubic nonlinearality. We ignore the nonlinear interaction in this paper and only briefly discuss its effect in Sec.~\ref{sec:results}.

We study the phonon thermal conductance of benzene and alkane chains covalently
bonded with two $(6,0)$ SWCNTs via the amide group. This is relevant to a
recent experimental setup\cite{guo06}, where the SWCNT is oxidatively cutted,
and the cutting edges are covalently bonded with molecular chains via the amide
group.  To get the force constants of the system, we do two separate runs for
the center and the leads using Gaussian03. For the center, we include extra one
and a half periods of SWCNT at each side, which is terminated by hydrogen atoms
at the outer boundaries. The cutting edges may be terminated by H or COOH
group. We optimize the center at the level of b3lyp density functional method
using the 6-31G basis set. We first relax the structure by constraining the
leads to be coaxial to get an optimized distance between the two leads. Then we
fix the outer half period of the SWCNTs and the H atoms at both sides and allow
all other atoms to move freely. The optimized structure is used to get the
dynamical matrix.  The inset of Fig.~\ref{fig:coohh} shows the optimized
benzene structure terminated by the H and COOH groups at the cutting edges. For
the lead, it is desirable to use periodic boundary condition to compute the
dynamical matrix, while this is out of the ability of Gaussian03. So we
optimise $7$ SWCNT periods and extract the force constant from the central
period to minimize the finite size effect. To connect the molecule with the two
leads, we remove the outer fixed atoms in the molecule and connect the
remaining period of SWCNT with one semi-infinite SWCNT at each side. For the
coupling matrix between the center and the leads, we only include coupling
between one period of SWCNT atoms in the center and one period in the leads.
This is a good approximation so long as we include enough SWCNT atoms into the
center molecule. 

In the NEGF method as described in
Refs.~\onlinecite{wang:033408,wang-2007,wang-review-epjb,mingo06,yamamoto06}, thermal
conductance of the molecular junction can be calculated from the 
Landauer formula ($\hbar=1$ throughout the formula)
\begin{equation}
\label{eq:caroli}
\sigma = 
\frac{1}{2\pi} \int_0^\infty \!\!\!d\omega\, \omega\, T[\omega] 
\frac{\partial f(\omega)}{\partial T}
\end{equation}
with the transmission coefficient
\begin{eqnarray}
T[\omega] &=& 
{\rm Tr} \bigl\{ G^r \Gamma_L G^a \Gamma_R \bigr\}. 
\label{eq-eff-trans}
\end{eqnarray}
$f(\omega)$ is the Bose-Einstein distribution function.
The retarded Green's function $G^r$ is obtained from 
\begin{equation}
	G^r[\omega] = \left( (\omega+i0^+)^2-K^C-\Sigma^r_L[\omega]-\Sigma^r_R[\omega] \right)^{-1},
	\label{eq:dyson1}
\end{equation}
where the retarded self-energy of lead $\alpha$ is 
\begin{equation}
	\Sigma^r_\alpha[\omega] = V^{C\alpha}g^r_\alpha[\omega]V^{\alpha C},
	\label{eq:sel}
\end{equation}
and the lead surface Green's function $g^r_\alpha[\omega]$ can be calculated
from the generalized eigenvalue method\cite{wang-review-epjb}, e.g., 
\begin{equation}
g^r_R[\omega] = \left( (\omega+i0^+)^2-k^R_{11}-k^R_{01}F^+_R(1) \right)^{-1}.
	\label{eq:surface}
\end{equation}
$k^R_{11}$ and $k^R_{01}$ are the diagonal and off-diagonal parts of the
right lead dynamical matrix. Their sizes are determined by the degrees of
freedom $M$ in each period of the lead. The matrix $F^+_R(s)$
translates to the right the displacement in the
$n$th period to the $(n+s)$th period $u^+_R(n+s)=F^+_R(s)u^+_R(n)$. 
It is constructed from the eigen values and vectors of the generalized eigen value problem
\begin{equation}
\label{eq:eigen}
\left( \begin{array}{cc} (\omega+i0^+)^2I\!\!-\!\!k^R_{11} & -I \\
             k^R_{10} & 0 
           \end{array}\right) 
\left( \begin{array}{c} \epsilon \\
             \zeta 
           \end{array}\right) = \lambda
\left( \begin{array}{cc} k^R_{01} & 0 \\
             0 & I 
           \end{array}\right) 
\left( \begin{array}{c} \epsilon \\
             \zeta 
           \end{array}\right)
\end{equation}
as $F^+_R(s)=E^+_R\Lambda_+^s(E^+_R)^{-1}$.  Here $I$ is an $M\! \times\! M$
identity matrix. The diagonal matrix $\Lambda_+^s$ consists of all the eigen
values $|\lambda_+^s| < 1$, and $E^+_R$ the corresponding eigen vectors $E^+_R
= (\varepsilon^+_1,\varepsilon^+_2,\cdots,\varepsilon^+_{M'})$. Note that $M'$
may be less than $M$, in which case the matrix inverse becomes pseudo-inverse.
A similar left-translation matrix $F^{-}_R(s)$ can be constructed from
$\Lambda_{-}^s$ and $E_R^-$, which include all the eigen values $|\lambda_-^s|
> 1$ (excluding infinity) and the corresponding eigen vectors.

While the NEGF method is systematic and suitable to take into account the
nonlinear interaction, the transmission coefficient from it is the sum of
all the eigen modes from the leads. Thus it is difficult to analyse the
contribution from each mode. The mode-matching method provides another way
to calculate the transmission coefficient in the ballistic
limit\cite{ando91,khomyakov05,wang-review-epjb}. Single mode transmission and
mode mixing effect can be studied by this method. The transmission matrix is
given by
\begin{equation}
	t^{RL}_{mn} = \sqrt{\frac{v_{R,m}^+a_L}{v_{L,n}^+a_R}}\tau^{RL}_{mn},
	\label{eq:mm1}
\end{equation}
and
\begin{equation}
	\tau^{RL} = (E^+_R)^{-1}g^r_RV^{RC}G^rV^{CL}g^r_Lk_{10}^L(F^{+}_L(-1)-F^{-}_L(-1))E^+_L,
	\label{eq:mm2}
\end{equation}
where $v^+_{\alpha,m}$ is the group velocity of the $m$th right propagating
mode for the lead $\alpha$. While the mode indices $m$ and $n$ are only for
propagating modes, the matrices $E^{\pm}_\alpha$ include all the propagating
and evanescent modes. The total transmission coefficient as in the Landauer
formula is $T[\omega] = {\rm Tr} \bigl\{ (t^{RL})^\dagger t^{RL} \bigr\}$.
Similar relations hold for waves incidented from the right lead and transmitted
to the left $t^{LR}_{nm}$. 

The NEGF and the mode-matching method are exactly equivalent in the ballistic
case as show in Ref.~\onlinecite{khomyakov05}. All the matrices needed by
Eqs.~(\ref{eq:mm1}--\ref{eq:mm2}) can be obtained by solving the generalized
eigen value problem Eq.~(\ref{eq:eigen}). It is interesting to note that by
doing a singular value decomposition on the transmission matrix $t$ we can get
the transmission eigenchannel information without any other efforts, for which
different methods have been developed in the electronic transport
literature\cite{inglesfield:155120}.

\section{Numerical results and discussion}
\label{sec:results}
\begin{figure} 
\includegraphics[scale=0.6]{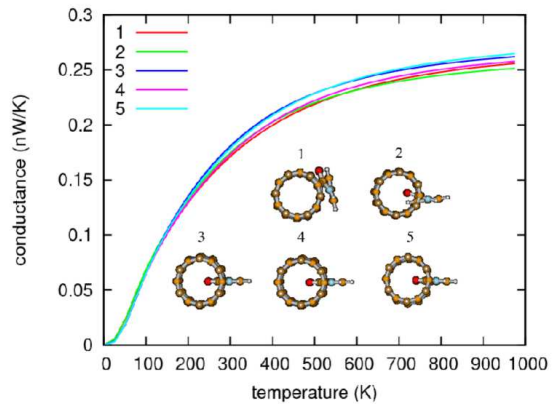}
\caption{\label{fig:confs} Ballistic thermal conductance of the benzene junction terminated by H atoms at different configurations. $3$ is the full relaxed structure. $1-2$ are in the compression state, and $3-4$ are in the stretching state.}
\end{figure}

We now present our numerical results. We begin with the simplest case where
there is only one benzene ring at the center. We compare the thermal
conductance of the molecular junction at different compression or stretching
configurations by changing the distance between the two leads. The purpose of
this study is twofold. Firstly, we want to study the effect of the compression
or stretching on the thermal conductance. Secondly, our optimization process
discussed in Sec.~\ref{sec:theory} can not ensure that we have reached the
lowest energy configuration while keeping the two SWCNTs to be coaxial. This is because we have fixed the position of the outer C and H atoms at both sides. If the thermal conductance is not sensitive to the
distance between the two leads, our results make sense even if we do not find the
lowest energy configuration. As we can see in Fig.~\ref{fig:confs}, this is
indeed the case. In Fig.~\ref{fig:confs}, from $1$ to $5$ the distance between the two carbon
atoms connecting to the amide groups is $10.07$, $10.21$, $10.38$, $10.51$,
$10.64$ \AA, respectively. The inset shows the atom configuration of the five
cases. In the compression states ($10.07$ and $10.21$ \AA), the relative position between the benzene ring and the SWCNT leads changes compared to the full relaxed ($10.38$ \AA) and stretching ($10.51$ and $10.64$ \AA) states (inset of Fig.~\ref{fig:confs}). Although the atom configurations change a lot, the ballistic thermal
conductance of the molecular junction only changes slightly in all the
temperature range studied here. So we can conclude that the ballistic thermal
conductance of the molecular junction is not sensitive to small compression or
stretching of the molecule.

\begin{figure}
\includegraphics[scale=0.6]{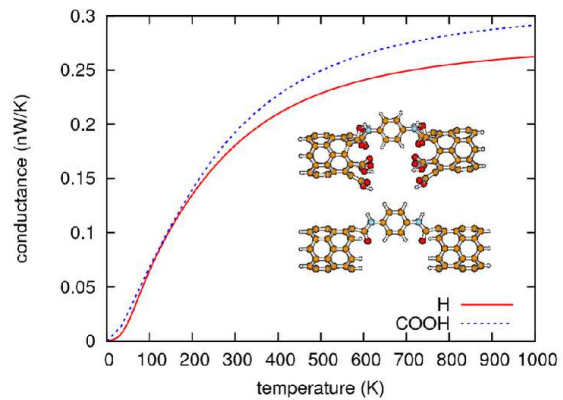}
\caption{\label{fig:coohh} Ballistic thermal conductance of the benzene junctions terminated by H and COOH group.}
\end{figure}

In a related experiment\cite{guo06}, the cutting edge carbon atoms are
expected to be saturated by the COOH group, not the H group shown in
Fig.~\ref{fig:confs}. In Ref.~\onlinecite{renw07}, the authors show that the
electron transmission is largely affected by the terminating groups. It is also
interesting to know how the terminating group affects the quantum thermal
conductance. We still use the single benzene structure to study this problem.
Figure~\ref{fig:coohh} shows the thermal conductance of the two configurations.
The full relaxed structures are shown in the inset. The electrical conductance
is mainly determined by the energy channel near the chemical potential, while
the thermal conductance is jointly contributed by many phonon modes. The
terminating group only has a large influence on the high energy (short
wavelength) optical phonon modes.  The low energy (long wavelength) phonon
modes are not sensitive to the local environment at the cutting edges.  As a
result, the thermal conductance only shows quantitative difference, which
becomes larger at high temperatures.

\begin{figure}
\includegraphics[scale=1.0]{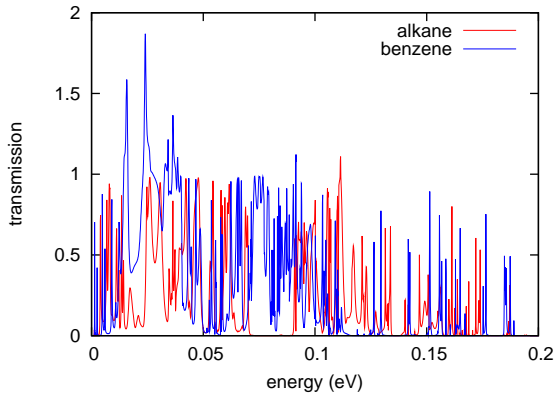}
\caption{\label{fig:3} Phonon transmission probability as a function of energy for the benzene and alkane chains.}
\end{figure}

\begin{figure}
\includegraphics[scale=1.0]{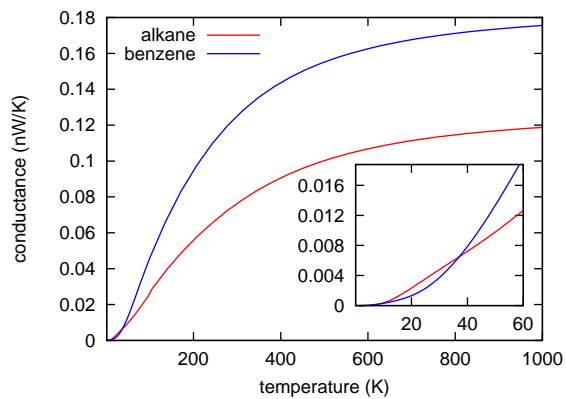}
\caption{\label{fig:4} Ballistic thermal conductance as a function of temperature for the benzene and alkane chains. The inset shows the crossing point of the thermal conductance at about $38$ K.}
\end{figure}

We now compare the conductance of the benzene rings and that of the alkane
chains, with the SWCNTs terminated by H atoms. We expect that the COOH
terminating structures show similar behavior. To ensure the two molecules have
comparable length, we include two benzene rings for the benzene structure and
$8$ CH$_2$ groups for the alkane chain. The total transmission probability and
the thermal conductance are shown in Figs.~\ref{fig:3} and \ref{fig:4},
respectively. Since the nonlinear interaction is not considered here, phonon
modes with different energies are independent of each other. The transmission
probability is nonzero only in the overlapped energy range of the SWCNT leads
and the center molecule.  Strong coupling with the leads gives rise to wide
broadening and strong shifting of the discrete molecule phonon energy. It is
hard to make a one-to-one correspondence between the transmission peaks and the
isolated molecule phonon eigen frequencies. This is especially true for low
energy phonon modes, which have relatively large spatial extent. The high
energy modes are highly localized and show sharp peaks in the transmission
spectrum. Due to the highly localized nature and low Bose-Einstein weighting,
these modes can hardly transfer energy. There is a wide zero transmission gap
around $0.07$ eV for the alkane chain, which is a characteristic of the alkane
chain phonon spectrum\cite{segal:6840}.

\begin{figure}
\includegraphics[scale=0.8]{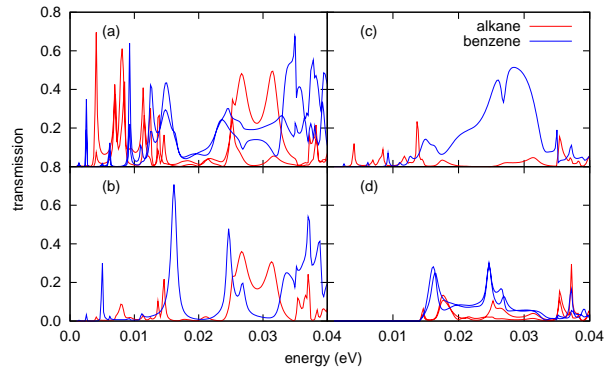}
\caption{\label{fig:5} Transmission probability of different phonon branches: (a) TA modes, (b) LA mode, (c) twisting mode, (d) lowest optical modes.}
\end{figure}

To gain further insight into the transmission spectrum, we also study the
single mode transmission using the mode-matching method. Figure~\ref{fig:5}
shows the transmission of several important branches in SWCNT phonon spectrum: Two transverse acoustic (TA) phonon
modes, one longitudinal acoustic (LA) mode, one twisting mode and the lowest two degenerate optical modes. These
low frequency modes are expected to contribute much to the thermal
conductance. The degenerate phonon modes of SWCNTs show different transmission
in both cases. This is consistent with the fact that the junction structure
destroys the degeneracy in transverse directions. At the energy range below
$0.01$ eV, the main contribution comes from the TA
modes of the SWCNT. The alkane chain shows larger transmission than the
benzene chain in this energy range. Above $0.01$ eV, the benzene chain shows
larger transmission in most cases.  The twisting mode shows the largest
difference in two structures. While it can hardly be transmitted by the
alkane chain, it has a large contribution in the benzene chain.

The thermal conductance of the molecular junction depends not only on the
transmission coefficient, but also on the phonon occupation number, which is
reflected as the Bose-Einstein distribution in the Landauer formula. According
the analysis of the transmission spectrum, we may expect that the alkane chain
has larger thermal conductance at low temperatures, while at higher
temperatures the order reverses. This is confirmed in Fig.~\ref{fig:4}. The
crossing temperature is about $38$ K. The room temperature thermal conductance
is about $0.075$ nW/K for alkane chains and $0.125$ nW/K for benzene chains.
In a recent experiment\cite{wangzh07}, the thermal conductance of alkane chain
is found to be smaller than our theoretical value. This difference comes from
the effect of the leads. We are using SWCNTs here, while in the experiment it
is the bulk gold connected with the alkane chain via the sufur atom. At least
two factors from the leads may account for this decrease. The first is the
smaller phonon spectrum overlap between gold and alkane chain, and the second
is the weaker coupling between them.

Some comments are worthwhile. The nonlinear interaction may change the
transmission spectrum, and lead to a decrease of the thermal conductance.  A
perturbative analysis of the single benzene structure shows that the room
temperature thermal conductance drops about $30\%$ of the ballistic
value if we include the cubic force constant calculated from Gaussion03.
Mean-filed approximation in the NEGF formalism only works well for simple
structures with relatively weak nonlinear interaction\cite{wang-review-epjb}. It
fails to converge in the self-consistent iteration in present case. So it is
still a challenge to find a good approximation for the nonlinear
self-energies in NEGF method.

\section{Conclusions} In this paper, we introduce a straightforward method to
calculate the phononic thermal conductance of molecular junctions in the
ballistic regime from first-principles. The force constant matrices are
obtained from Gaussian03 quantum chemistry software. The phonon transmission
and thermal conductance are calculated using the NEGF or mode-matching method.
Further information can be obtained from the transmission spectrum of each
single mode. Using this method we show that the ballistic thermal conductance
of the benzene ring amide linked with SWCNTs is not sensitive to the distance
between the two SWCNTs. The benzene rings show larger thermal conductance
than the alkane chains. This method is general, and can be easily applied to
other material systems.

\begin{acknowledgments}
The authors thank Nan Zeng for discussions. JTL is grateful to the hospitality
of Prof. J. C. Cao at Shanghai Institute of Microsystem and Information
Technology, where this paper was finished. This work was supported in part
by a Faculty Research Grant (R-144-000-173-101/112) of National University of
Singapore.
\end{acknowledgments}

\bibliography{molecule}

\end{document}